\documentclass[12pt]{article}
\pdfoutput=1

\usepackage{authblk}
\usepackage{geometry}
\usepackage{hyperref}
\usepackage{amsmath}
\usepackage{enumitem}
\usepackage{graphicx}

\title{Modeling Speculative Trading Patterns in Token Markets: An Agent-Based Analysis with TokenLab}

\author[1]{Mengjue Wang}
\author[2,3,4]{Stylianos Kampakis}

\affil[1]{University College London}
\affil[2]{UCL Centre for Blockchain Technologies}
\affil[3]{London Business School}
\affil[4]{Tesseract Academy}

\date{}

\begin{document}

\maketitle

\begin{abstract}
This paper demonstrates the application of Tokenlab, an agent-based modeling framework, to analyze price dynamics and speculative behavior in token economics. Tokenlab simplifies the simulation of complex token systems by decomposing them into discrete agent interactions governed by fundamental behavioral rules. The framework's primary innovation lies in its ability to model diverse speculative strategies and their impact on token price evolution.

Through a novel controller mechanism, Tokenlab can simulate multiple speculator archetypes and their interactions, providing insights into market sentiment and price formation. This approach enables the systematic study of how varying levels of speculative activity and strategies at different market stages influence token price dynamics. Our findings contribute to the understanding of speculative behavior in token markets and offer a quantitative framework for analyzing market heat indicators.
\end{abstract}

\section{Introduction}

Utility tokens are one of the most prevalent type of tokens, primarily used to provide access to products or services within a specific platform or ecosystem, or, saying, a medium of exchange. However, due to the inherent volatility of token prices, they inevitably become objects of investment or speculation. Clearly, this reality introduces a certain level of complexity to the valuation of utility tokens.

Three representative valuation methods frequently mentioned for utility tokens, the supply-demand model, the network value model, and the income approach, reflect three different perspectives on token value: the supply-demand model views tokens as products in a transaction, the network value model sees token price as a representation of network value, and the income approach treats tokens as securities, with their value reflecting a company’s valuation.

Beyond these methods, a tokenomics model based on the Quantity Theory of Money is also an important valuation approach. In this theory, price describes the overall price level of goods and services within an economy, essentially reflecting the relationship between the quantity of money and the demand for it in economic activity. Compared to traditional economies, blockchain-based token ecosystems are more easily studied using the Quantity Theory of Money because the supply and transactions of tokens can be fully recorded. How token issuance, circulation, locking mechanisms, and burning mechanisms can affect the supply-demand relationship between tokens, fiat currencies, and services within the ecosystem can be more clearly described.

The common problem with these valuation methods is that the theoretical price of tokens is usually very different from the actual price, and an important reason for this is that there is a great deal of speculative behaviour in the token system. In the case of roughly stable supply and demand over the long term, the contribution of speculator behaviour to token price volatility is even dominant. However, speculative strategies are always adjusting with market price movements and the number of speculators is constantly changing, modelling the speculative behaviour of the system as a whole is difficult, and a game-theory based agent model may be a better choice.

Tokenlab is an ABM-based token simulation framework that enables token price simulation. By adding different controllers, Tokenlab can control token issuance and burning, agent behaviour, and other token ecosystem parameters, which in turn enables the simulation of token prices. The contribution of this paper is the development of a new supply controller for speculative behaviour within Tokenlab, which simulates token price fluctuations and market speculation at different times. For simplicity, this paper first constructs a simple model that simulates the behaviour of speculators and their impact on token prices by setting the proportion of speculative trades in the overall trading and defining specific speculative buying and selling strategies. The aim of this simulation is not to precisely match daily price movements, but rather to attempt to explain the phasing of token price trends through the phasing of speculative activity. This approach provides a new perspective for understanding and explaining the intensity and attributes of speculative activity behind market price.

\section{Background Information}

\subsection{Token Taxonomy}

Tokens in the blockchain ecosystem generally fall into three distinct categories: utility tokens, security tokens, and cryptocurrencies \cite{oliveira2018}. Utility tokens serve as access mechanisms within specific decentralized applications (DApps) or blockchain platforms, functioning similarly to prepaid vouchers. Initially distributed through Initial Coin Offerings (ICOs), these tokens derive their primary value from their utility within their respective platforms. For example, Ethereum's \$ETH enables users to pay network fees and deploy smart contracts, while exchange tokens like Binance's \$BNB and Uniswap's \$UNI facilitate trading and fee payments on their respective platforms.

In contrast, security tokens represent traditional financial instruments (equity, debt, or asset ownership) in digital form, while cryptocurrencies like Bitcoin serve as general-purpose payment and value storage mechanisms across the broader blockchain ecosystem \cite{diangelo2020}.

This paper focuses on utility tokens, the most prevalent token category in blockchain ecosystems. Two key features distinguish utility tokens and warrant particular attention: their value proposition stems from platform utility rather than traditional financial rights—users purchase them to access services, often at preferential rates compared to other payment methods; Despite their intended function as service facilitators, utility tokens frequently become targets for speculation due to their price volatility, creating a dual nature that complicates their valuation.

\subsection{Tokenomics and Token Value}

Tokenomics encompasses three distinct domains: token issuance mechanism design, cryptoeconomics, and the economic incentive structures within token systems \cite{kampakis2022}. The method in this paper primarily utilises the analysis on the first aspect—the economic architecture of tokens, including their issuance, distribution, circulation, incentives, and governance \cite{kampakis2018}. Blockchain technology uniquely enables tokenomics research by providing comprehensive transaction records and transparent smart contract implementations, offering advantages over traditional economic analysis.

The valuation methodology for tokens varies by type. While security tokens can leverage established financial valuation methods due to their direct relationship with underlying assets, utility tokens present more complex valuation challenges. Utility token prices reflect a combination of platform success and service demand, token supply-demand dynamics, and ecosystem growth potential. These factors make utility token prices essentially a proxy for their ecosystem's overall value, distinguishing them from traditional assets and requiring specialized valuation approaches.

\subsection{Utility Token Valuation Methods and Limitation}

Two significant approaches to token valuation are the income approach and the network model, each offering distinct perspectives but facing notable limitations.

The income approach applies traditional Discounted Cash Flow (DCF) analysis to token valuation by assessing the future cash flows generated by the associated protocols and services \cite{xu2022}. While this method provides clear numerical benchmarks through comparisons with traditional financial ratios (P/E and P/B), it oversimplifies token value by assuming it depends solely on protocol service demand. This approach fails to capture the token's broader ecosystem value, including its roles as a medium of exchange, store of value, and investment vehicle.

The network model, based on Metcalfe's Law, posits that a network's value grows proportionally to the square of its user count \cite{peterson2018}. Applied to cryptocurrencies, this model uses wallet addresses as nodes to estimate network value. While studies suggest this method effectively explains medium to long-term price movements, its primary limitation lies in the unreliable correlation between wallet addresses and actual users, as single users often maintain multiple wallets for different purposes.

Both methods contribute valuable insights to token valuation but fall short of capturing the full complexity of token price dynamics.

\subsection{Quantity Theory of Money in Crypto Valuation}

The price formation mechanism of this paper's method is based on the Quantity Theory of Money (QTM). The QTM is a classical economic theory that suggests a direct relationship between the quantity of money in an economy and the price level of goods and services \cite{friedman1989}. The Equation of Exchange (EoE) is the mathematical representation of the QTM and is expressed as:

\begin{equation}
M \times V = P \times Q
\end{equation}

Where:
\begin{itemize}
    \item M is the money supply,
    \item V is the velocity of money (the average number of times a unit of money changes hands in a given period),
    \item P is the price level,
    \item Q is the volume of transactions or the real output in an economy.
\end{itemize}

Initially, it was assumed that money velocity was stable or at least predictable in the short run. This implied that any change in money supply would directly affect price level, which formed the basis for many monetary policies, where the real gross domestic product can be represented as Q in the EoE.

Despite its theoretical soundness, the QTM faces significant practical challenges in traditional monetary economics \cite{congdon2024}. These challenges stem from two primary sources of ambiguity. The first relates to money supply measurement: central bank base money (M0) proves inadequate for the exchange equation, while broader measures like M1 or M2 introduce definitional and standardization complexities. The second concerns transaction volume measurement, where economists must choose between using national income or actual transaction volume—each choice leading to distinct interpretations of the equation.

Money supply and transaction volume function as system state parameters, but traditional economic systems, similar to Web 2.0, lack clear state visibility or tracking mechanisms. This inherent statelessness, combined with limited data accessibility, has historically impeded the practical application of QTM in conventional economic analysis.

In contrast, the blockchain environment offers a unique advantage for applying QTM due to its inherent statefulness \cite{voshmgir2020}. Unlike traditional systems, blockchain networks maintain complete, immutable records of token supply and all transactions, enabling precise analysis of economic ecosystem dynamics. Vitalik Buterin (2017) pioneered the adaptation of QTM to token economics by redefining its key variables: token supply (M), token velocity (V), token-denominated price of ecosystem goods and services (P), and daily transaction value (T). This framework introduces two critical relationships: token price in fiat currency (C) as the inverse of P, and average token holding time (H) as the inverse of velocity. These relationships yield a simplified equation:

\begin{equation}
M \times C = T \times H
\end{equation}

This formulation reveals that token market capitalization (left side) equals the product of transaction volume and holding time (right side). Consequently, token price can increase through various mechanisms: growing transaction volume from increased user adoption, longer token holding periods, or supply constraints through mechanisms like staking or burning.

The blockchain's comprehensive state tracking makes this application of QTM particularly effective for analyzing and forecasting token values. For example, Pazos \cite{pazos2018} demonstrates a practical application of QTM to utility token valuation, starting by determining the monetary system requirements for specific transaction volumes. The approach combines economic growth modeling with Net Present Value (NPV) calculations to examine how token prices relate to economic expansion through EoE. The resulting valuation model identifies three key relationships: token value increases proportionally with resource prices and addressable market size while decreasing with higher token velocity.

\section{Methodology}

\subsection{Tokenlab as a Crypto Simulation Tool}

\begin{figure}[ht]
    \centering
    \includegraphics[width=0.7\textwidth]{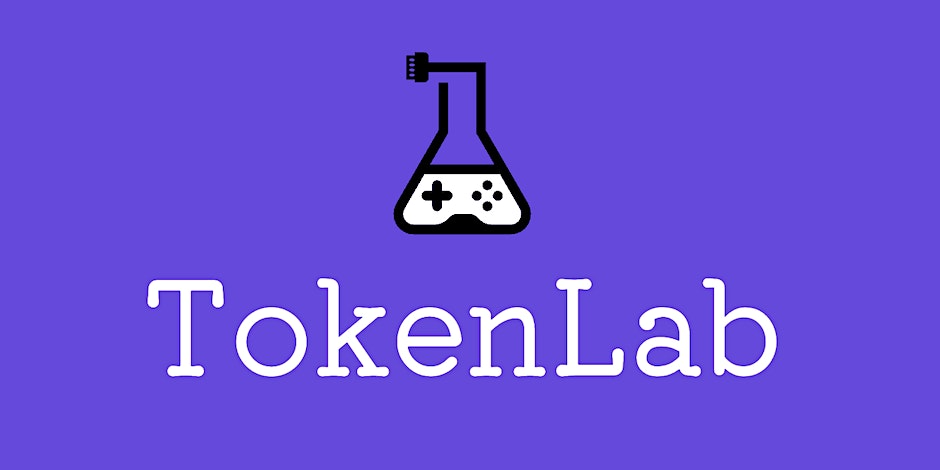}
    \caption{TokenLab Logo}
    \label{fig:tokenlab_logo}
\end{figure}

Tokenlab is an agent-based token economy simulation library that serves as a powerful tool for modelling, analysing, and auditing token economies. With Tokenlab, users can focus more on the parameters and rules of the token economy without delving deeply into the algorithmic implementations. Tokenlab is characterised by its modularity, clarity, abstract intermediary layers, focus on the economic aspects, and flexibility \cite{kampakis2024_tokenlab}. It allows for the simulation of various token systems through its diverse abstractions of agent pools, token economies, and controllers. Tokenlab is available at: \url{https://github.com/stelios12312312/TokenLab}.

The main configuration options are as follows:
\begin{itemize}
    \item \textbf{Agent Pool}: Tokenlab can simulate different types of agent pools, such as simple agent pools, buyback agent pools, or staking agent pools.
    \item \textbf{Supply}: The supply can be set as a constant or can be read from data. It can also simulate behaviours like bonding curves, cliff vesting, randomness, token burning, selling, or cross-period freezing.
    \item \textbf{User Growth}: The growth can be assumed to be stable, or modelled using linear, exponential, logistic, or other classic models, or even random numbers.
    \item \textbf{Transactions}: Transactions can be set as a constant, read from data, or generated randomly according to specific distributions or trends. Transaction volumes can be set proportional to market capitalisation, or more complex rules can be implemented in multi-token systems.
    \item \textbf{Holding Time}: Holding time can be read from data, assigned randomly, or calculated based on transaction volumes and market prices.
    \item \textbf{Pricing}: Token pricing mechanisms can be based on the equation of exchange, bonding curves, or issuance curves.
    \item \textbf{Other Configurations}: Staking rules, treasury rules, etc.
\end{itemize}

The token economy simulation in Tokenlab updates the state of the token economy iteratively. Each iteration represents a time period during which various parameters of the token economy, such as the number of users, transaction volumes, and token prices, are updated. Multiple sampling runs can be conducted to mitigate the effects of different random states.

\subsection{Explaining Price Volatility through the Modelling of Speculation}

The QTM posits that token price determination depends on both standard trading metrics (supply and volume) and token velocity (or holding time). This consideration of velocity is crucial because token holders exhibit fundamentally different holding patterns based on their purposes. While holding time appears as a macro-level average, it masks distinct behavioral patterns: service users employ tokens briefly as transaction media, while speculators maintain longer positions awaiting favorable price points.

This behavioral dichotomy between users and speculators provides a novel framework for analyzing token price formation. Price fluctuations often reflect cyclical shifts in the user-to-speculator ratio, helping distinguish between service-driven and speculative trading volumes.

To analyze these dynamics, Tokenlab implements a new supply controller modeling speculative behavior. The controller operates on the following principles:
\begin{itemize}
    \item Each iteration allocates a portion of trading volume to speculative activity.
    \item Speculative buying reduces market token supply temporarily.
    \item The controller tracks take-profit and stop-loss thresholds for all previous speculative transactions.
    \item When prices breach these thresholds (either above take-profit or below stop-loss), speculators sell tokens, increasing market supply.
    \item Net supply changes from buying and selling activities feed into the broader token economy, influencing price formation.
\end{itemize}

The simulation utilizes actual token supply and trading volumes while maintaining fixed reference holding times. By modulating the speculative trading proportion, the model reveals relationships between token prices and speculative activity across different market phases. Importantly, the model separates user and speculator behaviors through distinct mechanisms—speculative activity directly affects token supply, while holding time parameters capture only service users' behavior. This separation prevents parameter correlation and enables more accurate ecosystem analysis.

\subsection{Tokenlab Configuration for Modelling Price Volatility}

This analysis selects \$LINK as a case study, examining data from January 2020 through July 2024—a period chosen for its distinct price phase patterns. The transaction data is sourced from CoinGecko.com \cite{coingecko2024}. The study employs actual transaction volume, supply data, and initial price information. While specific token holding time parameters are derived from historical data, their exact values are less critical than establishing a baseline that aligns speculative supply pools with actual token price movements across different market phases.

Based on \$LINK's price characteristics across different periods, the analysis models five distinct speculator archetypes:

\begin{enumerate}
    \item \textbf{Limited Short-term Trading (Low Volume)}
    \begin{itemize}
        \item Speculative Trading Proportion: 30\%
        \item Take Profit: 120\% of Purchase Price
        \item Stop Loss: 80\%
        \item Reflects minimal speculative participation with short-term profit focus
    \end{itemize}
    \item \textbf{Massive Short-term Trading (High Volume)}
    \begin{itemize}
        \item Speculative Trading Proportion: 70\%
        \item Take Profit: 120\% of Purchase Price
        \item Stop Loss: 80\%
        \item Represents dominant speculative activity with short-term profit focus
    \end{itemize}
    \item \textbf{Balanced Speculative Trading}
    \begin{itemize}
        \item Speculative Trading Proportion: 50\%
        \item Take Profit: 180\% of Purchase Price
        \item Stop Loss: 40\%
        \item Accommodates both price-sensitive traders and longer-term holders
    \end{itemize}
    \item \textbf{Limited Patient Capital (Low Volume)}
    \begin{itemize}
        \item Speculative Trading Proportion: 30\%
        \item Take Profit: 240\% of Purchase Price
        \item Stop Loss: 25\%
        \item Models investors with higher risk tolerance and longer time horizons
    \end{itemize}
    \item \textbf{Massive Patient Capital (High Volume)}
    \begin{itemize}
        \item Speculative Trading Proportion: 70\%
        \item Take Profit: 240\% of Purchase Price
        \item Stop Loss: 25\%
        \item Represents substantial speculative participation with long-term holding focus
    \end{itemize}
\end{enumerate}

The simulation spans the duration of available transaction records, with 30 iterations conducted to ensure statistical reliability. Each iteration implements the previously defined five speculator types through distinct supply controllers, generating comparative price curves against the actual token price. The analysis employs a deviation metric:

\[
\text{Percentage Deviation} = \frac{\text{Simulated Price}}{\text{Actual Price}} - 1
\]

Given that daily transaction volumes and prices often exhibit significant volatility, the raw daily deviation calculations produce noisy results that complicate interpretation. To enhance clarity, the analysis applies a 30-day moving average to the deviation values, producing smoothed trend lines that facilitate easier comparison across different simulations. For each distinct market phase, the study calculates the mean percentage deviation between simulated and actual prices, enabling identification of which speculator archetype most accurately characterizes the dominant trading behavior during that period. This approach allows for a systematic analysis of how speculative behavior patterns evolve across different market phases.

\section{Results}

The study of \$LINK token price dynamics reveals distinctive patterns across three market phases: upward (pre-500th iteration), downward (500th-875th iterations), and stable (post-875th iteration). Through comparative analysis between simulated and actual prices, several critical insights emerge about the relationship between speculative behavior and price formation.

\begin{figure}[ht]
    \centering
    \includegraphics[width=0.7\textwidth]{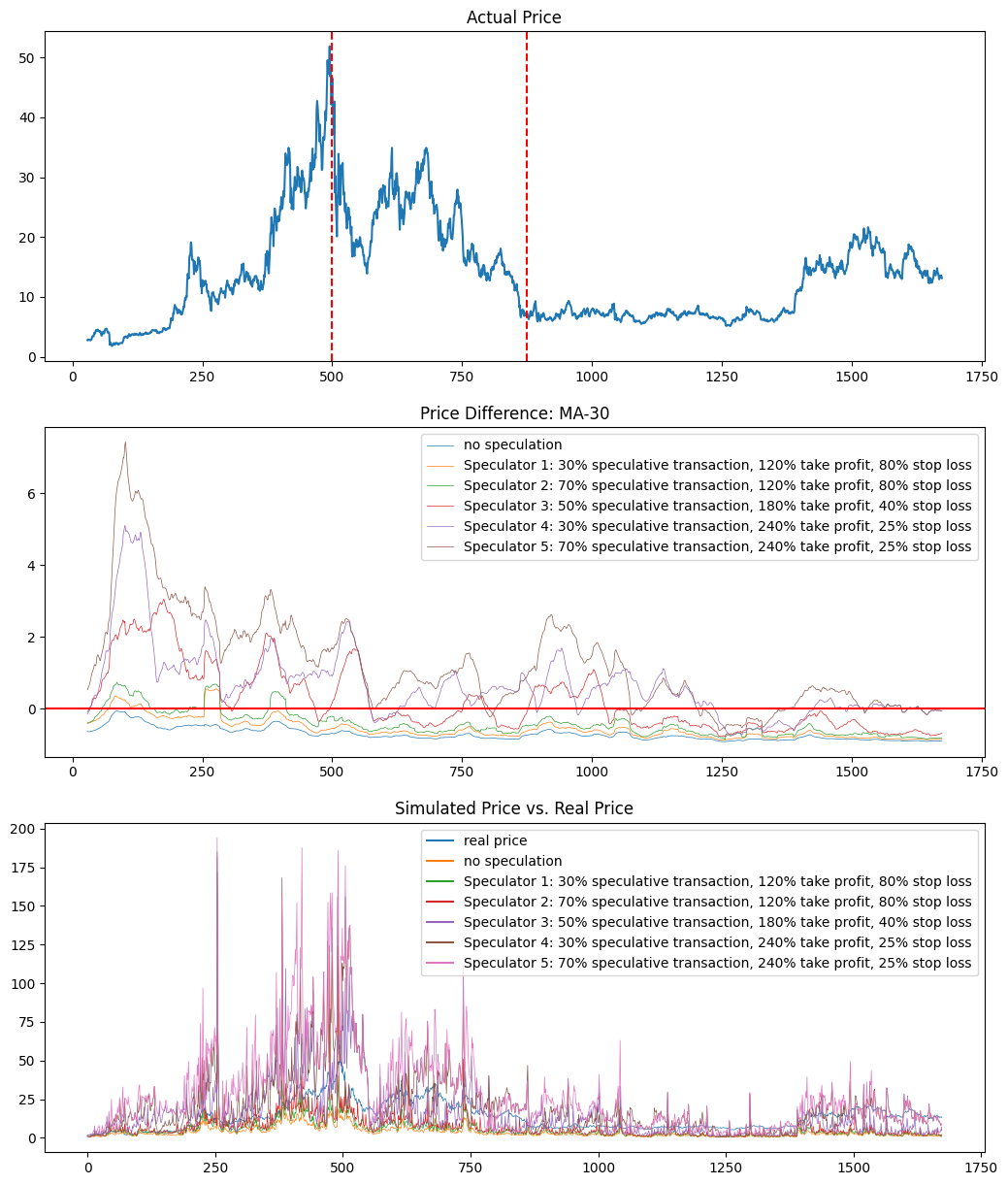}
    \caption{Simulated price dynamics and deviations. This figure illustrates the actual price trend, moving averages of price differences, and the impact of speculative behaviors on price simulations.}
    \label{fig:price_dynamics}
\end{figure}

As shown in Figure~\ref{fig:price_dynamics}, the price dynamics illustrate how speculative behaviors impact token prices over time. The deviations between simulated and actual prices are summarized in Table~\ref{tab:deviation_table}.

\begin{table}[ht]
    \centering
    \caption{Average percentage deviation: simulated prices vs. actual prices.}
    \label{tab:deviation_table}
    \begin{tabular}{|l|c|c|c|c|}
        \hline
        \textbf{Speculator} & \textbf{Before 500} & \textbf{500-875} & \textbf{After 875} & \textbf{After 1100} \\
        \hline
        No Speculator & -0.500583 & -0.778908 & -0.820052 & -0.856672 \\
        Speculator 1  & -0.213649 & -0.665884 & -0.733703 & -0.781814 \\
        Speculator 2  &  0.019085 & -0.550847 & -0.636397 & -0.701161 \\
        Speculator 3  &  1.200837 &  0.013630 & -0.307090 & -0.524879 \\
        Speculator 4  &  1.605689 &  0.563933 &  0.157062 &  0.049297 \\
        Speculator 5  &  2.919183 &  0.839544 &  0.452134 &  0.086036 \\
        \hline
    \end{tabular}
\end{table}

The simulations without speculator involvement consistently produced prices significantly below both actual market prices and speculator-inclusive simulations across all phases. This finding demonstrates that traditional supply-demand dynamics alone, based on transaction volume and token supply, cannot fully explain observed market prices. Furthermore, the results consistently show that higher proportions of speculative trading volume correlate with elevated price levels, even when maintaining identical buy-sell strategies.

Further analysis shows that within the three different phases, the simulated price closest to the actual price corresponds to a significant difference in the speculator hypothesis:
\begin{enumerate}
    \item \textbf{Phase 1 (pre-500th iteration)}: This upward momentum phase showed closest alignment with simulations featuring a high proportion of short-term speculators, yielding a +1.9\% deviation on a 30-day average. This suggests that price appreciation phases are primarily driven by concentrated short-term speculative activity, indicating a momentum-driven market where rapid trading and short holding periods dominate.
    \item \textbf{Phase 2 (500th-875th iterations)}: This market correction phase best matched simulations with a balanced mix of short-term and long-term speculators, showing a +1.4\% deviation on a 30-day average. This indicates that price declines involve complex interactions between different speculator types, representing a transitional phase where varying investment horizons and strategies coexist.
    \item \textbf{Phase 3 (post-875th iteration)}: The market stabilization phase initially showed high deviation (16\%) with simulations featuring a low proportion of long-term-focused speculators. However, refined analysis of the post-1100th iteration period revealed improved accuracy (-4.9\% deviation), suggesting that mature markets attract patient capital with reduced sensitivity to short-term fluctuations. This evolution indicates a transition toward a more stable, long-term-oriented investor base.
\end{enumerate}

\section{Conclusion and Future Directions}

This study's findings offer significant insights into the relationship between speculative behavior and token price dynamics, while demonstrating the effectiveness of agent-based modeling in token economics. The analysis reveals several key conclusions:

\subsection{Market Phase Characteristics and Speculator Behavior}

The simulation results demonstrate that different market phases attract distinct speculator archetypes. During the upward phase, short-term speculative activity with high trading volume dominates price formation (+1.9\% deviation), suggesting that rapid price appreciation attracts momentum-driven traders. The downward phase exhibits a more balanced speculator composition (+1.4\% deviation), indicating that market corrections involve a complex interplay between short-term traders and longer-term holders. The stable phase shows a preference for patient capital (-4.9\% deviation post-1100th iteration), suggesting that market maturity corresponds with longer investment horizons and reduced sensitivity to short-term fluctuations.

\subsection{Methodological Contributions}

The study validates Tokenlab's enhanced supply controller as an effective tool for modeling complex speculative dynamics. The controller's ability to simulate multiple speculator archetypes and their impact on token supply provides a novel framework for analyzing price formation mechanisms. The clear correlation between simulated and actual prices across different market phases demonstrates the model's robustness in capturing real-world market behavior.

\subsection{Practical Implications}

The findings have significant implications for token ecosystem analysis and investment strategy:
\begin{itemize}
    \item Market phase identification becomes possible through analysis of dominant speculator types.
    \item Token system comparisons can be made based on speculative composition.
    \item Investment decisions can be informed by understanding the current market's speculative dynamics.
\end{itemize}

\subsection{Future Research Directions}

While the current model provides valuable insights, several avenues for enhancement exist:
\begin{itemize}
    \item Integration of more diverse speculator archetypes with complex strategy combinations.
    \item Dynamic adjustment of speculator proportions based on market conditions.
    \item Incorporation of external factors influencing speculator behavior.
\end{itemize}

These findings not only contribute to the theoretical understanding of token economics but also provide practical tools for market analysis and investment decision-making. The demonstrated effectiveness of agent-based modeling in capturing market dynamics suggests this approach could become a standard tool in token ecosystem analysis.

\section*{Additional Resources}
\begin{itemize}
    \item Tesseract Academy. 2024. \url{https://tesseract.academy}
    \item TokenLab GitHub repository. \url{https://github.com/stelios12312312/TokenLab}
\end{itemize}

\end{document}